# Symmetry and planar chirality of a protein measured on an angular basis in a transmission electron microscope

A. H. Tavabi[1], P. Rosi[2], R. B. G. Ravelli[3], A. Gijsbers[3], E. Rotunno[2], T. Guner[4], Y. Zhang[3], A. Rocaglia[5], L. Belsito[5], G. Pozzi[1,6], D. Tibeau[1], G. Gazzadi[2], M. Ghosh[7], S. Frabboni[8], P. J. Peters[3], E. Karimi[4], P. Tiemeijer[8], R. E. Dunin-Borkowski[1], V. Grillo[2]

1. Ernst Ruska-Centre for Microscopy and Spectroscopy with Electrons and Peter Grünberg Institute, Forschungszentrum Jülich, 52425 Jülich, Germany
2. Centro S3, CNR - Istituto di Nanoscienze, 41125 Modena, Italy
3. Maastricht Multimodal Molecular Imaging Institute, Division of Nanoscopy, Maastricht University, Universiteitssingel 50, Maastricht 6229 ER, The Netherlands
4. Department of Physics, University of Ottawa, Ottawa, Ontario K1N 6N5, Canada
5. CNR - Istituto di Microelettronica e Microsistemi-CNR, 40129 Bologna, Italy
6. Department of Physics and Astronomy, University of Bologna, 40127 Bologna, Italy
7. Thermo Fisher Scientific, PO Box 80066, 5600 KA Eindhoven, the Netherlands
8. Dipartimento FIM, Universitá di Modena e Reggio Emilia, 41125 Modena, Italy

Corresponding authors:
Vincenzo Grillo vincenzo.grillo@nano.cnr.it
Raimond Ravelli : rbg.ravelli@maastrichtuniversity.nl

**In quantum mechanics, each conserved quantity (*e.g.*, energy, position, linear momentum and angular momentum) is associated with a Hermitian operator. Its expected value can then be determined by performing a measurement on the wavefunction[1-3]. In modern electron microscopy, one can select the initial and final states of the electron and the measurement basis by performing measurements of scattering processes[4,5]. For example, the orbital angular momentum (OAM) of an electron can be used to reveal the *n*-fold symmetry of a wavefunction scattered by a sample. Here, we introduce a new composite "planar chirality" operator that can be used to measure a "spiral-like" feature in a sample. This concept develops the concept of chirality to highlight a specific roto-scale symmetry. We show that planar chirality can be characterized using an electron OAM sorter to uncover the atomic structures of biomolecules in cryo electron microscopy[6], either in a stand-alone analysis for fast identification of protein structures or in the context of conventional cryo electron microscopy to produce faster and more detailed 3D reconstructions by solving upside-down orientation ambiguities.**

Symmetries are essential features of nature, providing insight into mechanisms of forces and conserved physical quantities[1,2,3]. As they are also essential features of biology and life in general, the determination of the symmetry of an object can provide important information about physical or chemical processes[4,5]. On a basic level, an object possesses discrete translational, rotational or parity symmetry when the Hamiltonian that describes its action is independent of the corresponding conjugate variables. This concept is known as the Noether theorem and the corresponding symmetry quantities are given by quantum operators, such as linear momentum, orbital angular momentum and the parity operator. When mirror symmetry is broken, a physical system becomes chiral[6]. Although chirality (which leads to phenomena such as birefringence and dichroism) can be defined, it is difficult to quantify "how chiral" an object is.

In a measurement instrument, it is essential to be able to probe physical quantities in the most direct way. The optimization of a measurement process, for example by increasing the number of quantities that are measured directly, is a new trend in electron microscopy[7,8,9,10,11,12]. However, the fact that real space representations still have a privileged role in quantum measurements lies at the foundation of quantum mechanics[13]. For example, measurements in an electron microscope are performed primarily in position (imaging), momentum (diffraction) and energy (electron energy-loss spectroscopy) spaces by making use of spatial dispersion through electron-optical elements. A recent addition to this list is the orbital angular momentum (OAM) sorter, which performs a wave transformation in log-polar coordinates and makes OAM decomposition visible on a detector[14]. For practical applications, an OAM sorter can be considered as an additional set of thin lenses, which break the cylindrical symmetry of the microscope[15,16,17,18]. OAM sorters based on synthetic electron holograms, electrostatic phase elements and structured magnetic elements have been demonstrated. In Cartesian coordinates (x,y), an OAM sorter performs the conformal transformation

$$(u_\rho, v_\theta) = \left(-s \cdot \log\left(\frac{\sqrt{x^2+y^2}}{b}\right), s \, arctan(y/x)\right),$$

where $s$ and $b$ are scaling parameters and $log(\cdot)$ and $arctan(\cdot)$ are logarithm and inverse tangent functions, respectively. After this transformation and ordinary diffraction, the quantity that is conjugate to $v_\theta$, i.e., OAM, is sorted and observed on a detector[19,20,21,22,23]. If the electron propagates in the $z$ direction, then the OAM quantum mechanical operator is $L_z = -i\hbar \frac{\partial}{\partial v_\theta}$, where $\hbar$ is the reduced Planck constant. In order to complete the set of observables, one also needs to define the radial basis. Here, we define the logarithmic radial momentum $P_\rho = -i\hbar \frac{\partial}{\partial u_\rho}$, which is a conjugate of $u_\rho$. This radial operator should not be confused with the radial operator for Laguerre-Gaussian (LG) modes, which is a second order derivative and depends on the azimuthal coordinate. In contrast, $L_z$ generates an in-plane rotation $R = exp(iL_z \delta v_\theta/\hbar)$. For an object with $m$-fold symmetry, rotation by $\delta v_\theta = 2\pi/m$ leads to the same wavefunction, meaning that the spectrum of $L_z$ is defined in multiples of $m$. The quantity $P_\rho$ has received much less attention. The transformation $T = exp(iP_\rho \delta u_\rho/\hbar)$ produces a translation of the wavefunction in $u_\rho$ space by an amount $\delta u_\rho$. The quantity $u_\rho$ is related to radial coordinates, with a translation $\delta u_\rho$ being related to radial scaling of the wavefunction as a result of the logarithmic factor.

We consider here a composition of the $R$ and $T$ symmetry operators in the form $S = R \otimes T$ to define a new planar symmetry that subtends an O(1)xU(1) group, as a function of the two parameters $\delta v_\theta, \delta u_\rho$. Waves that transform into themselves under this composite transformation have a logarithmic-spiral wavefront with $(\ell \, v_\theta + \kappa \, u_\rho) = const$, where $\ell$ and $\kappa$ are eigenvalues of $L_z$ and $P_\rho$. Logarithmic spirals are self-similar with the scaling $m = \frac{\ell}{\kappa}$ and represent a specific form of planar chirality and $m$-fold symmetry. The slope angle $\alpha = \arctan(m)$ is the angle formed by the spiral with every concentric radius. For example, in spiral galaxies, which are approximatively logarithmic spirals, this parameter has been found to be correlated with the size of the central black hole[24].

These considerations permit a new operator $\chi = L_z/P_\rho$, which we refer to as "spiral in-plane chirality", to be defined. The new variable $\chi$ measures how much a figure winds like a spiral, while its sign (clockwise/ counterclockwise) indicates the sign of the spiral. This concept is related to the in-plane chirality of an object, albeit without considering its three-dimensional structure. It is a pseudo-scalar quantity, since it is odd on performing a parity operation along $z$. (The angular momentum operator $L$ is a pseudovector, with $L_z$ changing sign upon $z$ inversion, while $P_\rho$ does not). The $z$ parity or mirror parity about a plane that contains $z$ can be transformed into a single mirror symmetry upon log-polar transformation and following diffraction. To be precise, mirror symmetry in the OAM-$P_\rho$ plane is transformed into mirror symmetry about the $L_z$=0 axis, while the direction of the symmetry plane is encoded in the phase. In this representation, it is intuitive to verify that an object that has at least one mirror symmetry plane is characterized by $<\chi>$=0. Since evolution through the focal region produces an inversion of the form (x,y) -> (-x,-y), it does not alter $\chi$. Unfortunately, as the additional phase that is introduced by defocus can alter the radial structure of the spiral, the quantity $\chi$ is only measured correctly in focus. Chirality is often related to the optical activity of an object and its differential response to an excitation such as light polarization[25]. Extrinsic chirality is related to the geometric form factor of the object. Here, we define the planar chirality of an object (or of a wave elastically scattered through that object) independently of the detail of the probe, such as its wavelength. The quantity $\chi$ then represents a "universal" symmetry of the object and is therefore scale invariant.

For a phase object, the value of $\chi$ is largely invariant to phase scaling by constant factor, for both the radial and the angular part. This is not true for the OAM and $P_\rho$ parts individually. This is a desirable property for a quantity that aims to measure only a geometrical factor, regardless of the strength of coupling with a probe.

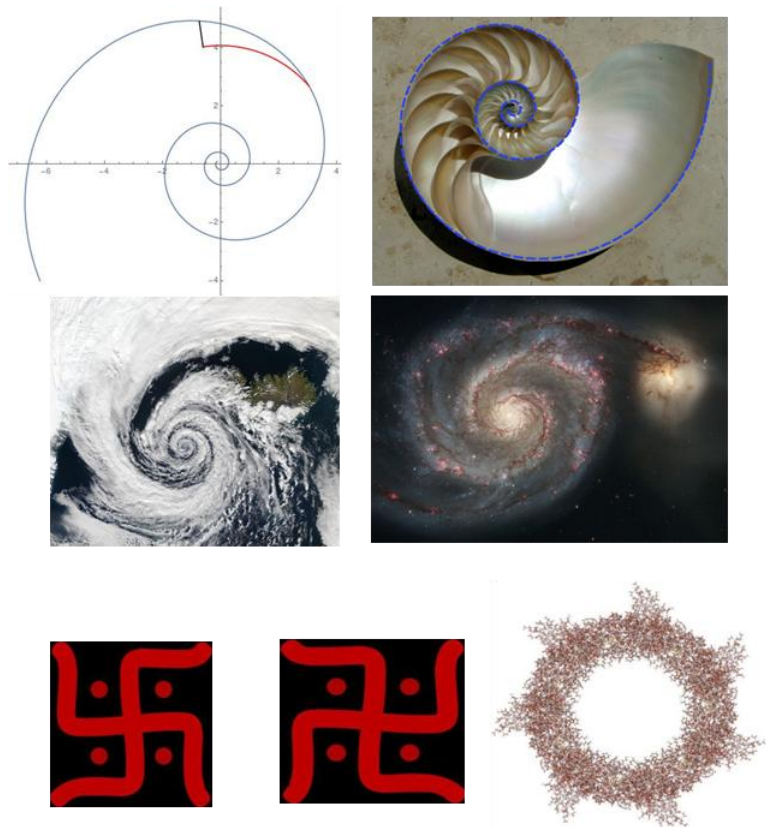

*Fig 1. Abstract logarithmic spiral and its realization in nature in shells, tornadoes and galaxies. Gammadions are well-known examples of 2D-chiral structures that are used in nano-photonic (e.g., plasmonic) studies. The lower right inset shows the structure of the* ESX-1 *secretion-associated protein B (EspB), which is the subject of the present study.*

Figure 1 shows examples of planar chirality in nature. Here, we concentrate on the ESX-1 secretion-associated protein B (EspB) from *Mycobacterium tuberculosis* oriented along its main symmetry axis[26]. Considering that real-space imaging is the main feature of an electron microscope, the most successful refinement of this idea would be its application in cryo electron microscopy for the analysis of dose-sensitive proteins, which typically involves imaging multiple instances of the same object in random orientations embedded in a layer of vitreous ice[27,28]. Although this approach can be used for 3D reconstruction of dose-sensitive objects such as proteins, it is not yet applicable at the level of a single protein. However, direct identificaton of the symmetry of an indiviudal protein would be of interest, both because ambiguity in up-down orientation can create ambiguity in 3D reconstruction and because it could allow the identification of specific characteristics of an object in a cell in the presence of other particles.

Elastic transmission electron microscopy (TEM) imaging of proteins results in a projection that does not inherently contain 3D information. However, in many cases the projected structure is not mirror symmetric about a plane, *i.e.*, it has "planar chirality". Here, we demonstrate that the use of an OAM sorter in the TEM allows the symmetries of a protein to be identified efficiently at a single-particle level within typical low-dose constraints. For simplicity, we consider a "virtual protein" produced by a computer-generated synthetic hologram[29]. The hologram is obtained by sculpting a thickness modulation in a thin membrane of SiN. This object is, to a good approximation, a phase object, *i.e.*, an object whose primary effect is to modulate the phase of the electron wave that traverses through it. The action of the hologram is similar to that of a real protein, *i.e.*, modulation of the phase of the incoming electron. Although its lateral scale is 100 times larger that that of a real protein, it is decreased by positioning the hologram in the illumination condenser aperture of the microscope. In terms of the electron wavefunction, it is therefore equivalent to studying a virtual protein.

An electrostatic OAM sorter, which has already been demonstrated to work for simple structures[7], was placed in the lower part of the electron microscope. The components of the OAM sorter must be centered carefully with respect to each other and the optical axis of the microscope. Figure 2 shows experimental results, which confirm that the sorter produces a log-polar conformal mapping of a wavefunction, which is then diffracted to form an OAM spectrum. Under ideal conditions, the OAM sorter produces a peak width with an OAM value of 1 in units of $\hbar$. In calibration experiments such as that shown in Fig. 2, it is now possible to reach a near-ideal OAM resolution of $\Delta\ell = 1.1$ through an appropriate redesign of the electrodes and the use of artificial intelligence for alignment of the system[30].

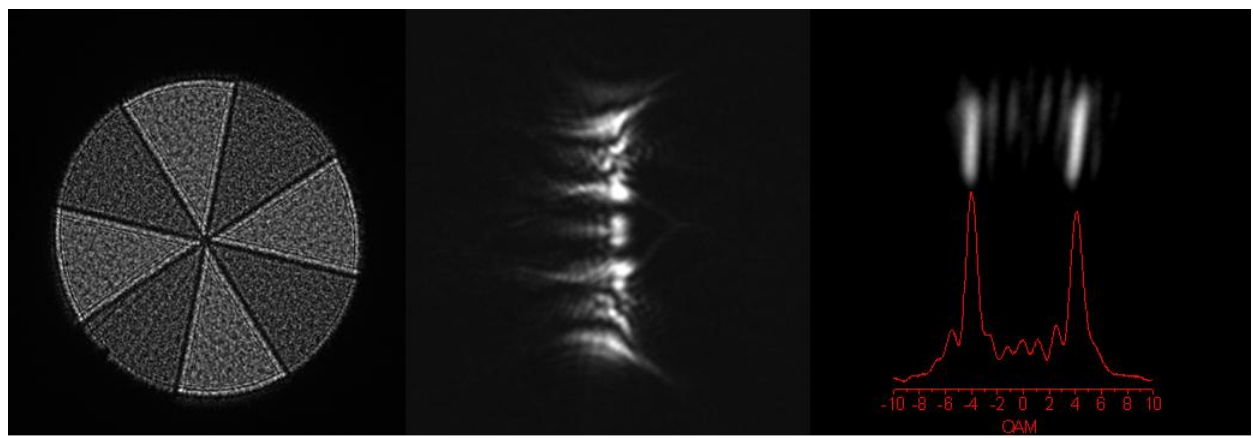

*Fig 2 OAM decomposition in a TEM for a non-chiral structure demonstrating 1.1$\hbar$ OAM resolution.*

Figure 3 shows experiments performed using holograms of virtual proteins corresponding to two opposite orientations, as well as for several different spiral beams. The second column shows real-space images recorded in focus. The third column shows experimental log-polar representations obtained using the sorter. The fourth column shows the OAM–$P_\rho$ representation, *i.e.*, the final outcome of sorter diffraction. The fifth column shows simulation results.

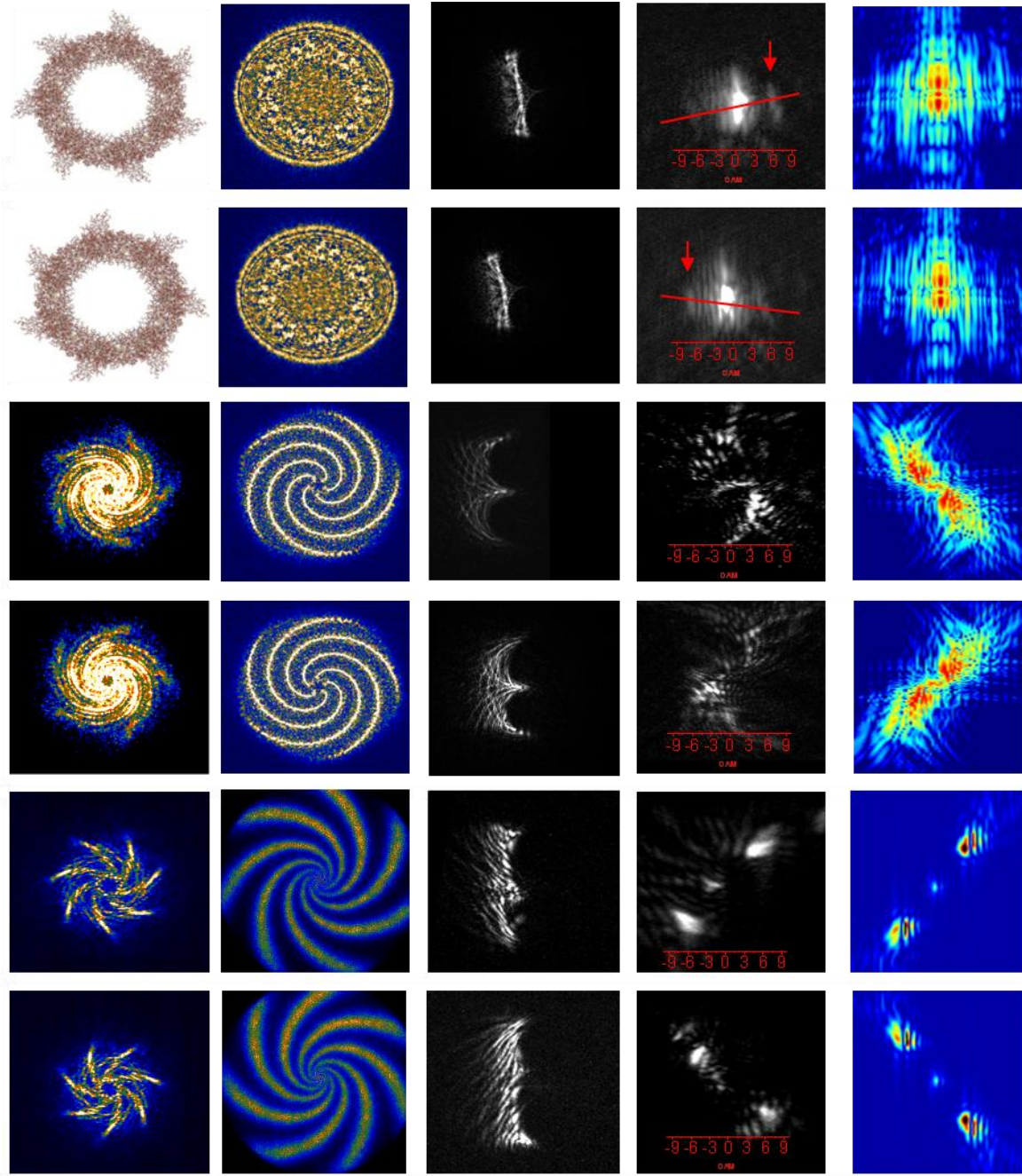

*Fig. 3 Experimental analysis of a virtual protein, an Archimedean and a logarithmic spiral experimentally analyzed by an OAM sorter, showing the resulting OAM-$P_\rho$ spectrum. The comparative analysis is performed on both orientations here shown in couple. See text for details. The SEM/TEM image of the object (or its structure are shown in column 1 and 2. The unwrapping of the wave is in column 3 and the experimental spectrum is in column 4. The relevant simulation is visible in column 5.*

The OAM sorter has the unique ability to evaluate the planar chirality that cannot be estracted by direct imaging. For example the 7-fold symmetry of the virtual proteins in Fig. 3 is not evident in in-focus images. Although it could be determined from out-of-focus images, they do not readily help to identify the clockwise or counterclockwise orientations of the "wings" of individual proteins as we will see in detail A full set of out-of-focus images is shown in the Supplementary Material.

Using the OAM sorter we found that the spiral-chirality defined above is $\langle \chi \rangle = +0.036$ and $\langle \chi \rangle = -0.028$ for the two protein cases shown in Fig. 3. As the change in sign was consistent in all experiments, it was possible to distinguish the two orientations based only on the sign of $\langle \chi \rangle$. For comparison, we obtained $\langle \chi \rangle = 0.005$ for the non-chiral structure shown in Fig. 2. This small deviation of $\langle \chi \rangle$ from zero is likely to be associated with uncertainty in the measurements and imperfections in alignment, hologram fabrication and the instrument. The absolute values of $\langle \chi \rangle$ for the two protein cases are relatively small, as the asymmetry is located only in their small external structures. The measured values are in good agreement with simulations, for which a range of $|\langle \chi \rangle| \in [0.020, 0.036]$ is obtained.

In the lower part of Fig. 3, Archimedean and logarithmic spirals, for which the chirality is relatively large, are considered. These spirals are characterized by wavefronts of the form $\ell\, v_\theta + k u_\rho = const$ and $\ell\, v_\theta + kr = const$, respectively. For the Archimedean spiral, the OAM-$P_\rho$ spectrum contains two staggered speckles with a left-right asymmetry. For the logarithmic spiral (lower two rows), the staggered points are more similar to dots, since log-spirals are eigenstates of the $L_z$ and $P_\rho$ operators (unlike vortex beams). For a clockwise spiral, the signs of OAM and $P_\rho$ are opposite, *i.e.*, they are either positive/negative or negative/positive. The extreme situation for the Archimedean spiral produces values of $\langle \chi \rangle = -0.157$ and $\langle \chi \rangle = 0.332$, which are larger than for the proteins. These values of $\langle \chi \rangle$ correspond to slope angles of -8° and 18°, respectively. However, for an Archimedean spiral the slope angle is not constant (resulting in blurring of the OAM–$P_\rho$ spectrum). At the periphery of the hologram, its value is approximately 17°. In the case of a log-spiral, the value of $\langle \chi \rangle$ is perfectly defined, since it is an eigenvalue of the $\chi$ operator and we find $\langle \chi \rangle = 0.55$ and $\langle \chi \rangle = -0.48$. The associated slope angles of 29° and 25° are very close to the actual value of 32° in the hologram.

It is worth noting that $\chi$ is a self-adjunct operator in Hilbert space and that its spectrum, based on the full complex wavefunction, can be calculated exactly in the OAM-$P_\rho$ representation as $[\chi, L_z]=[\chi, P_\rho]=0$. We analize here if the OAM soter analysis could be substituted by a simple digital analaysis. In facts we consider here the digital application of a log-polar transformation to an image of the protein. In this case, the phase information that is most relevant for the protein is only imaged out-of-focus, which changes the $P_\rho$ decomposition. For phase objects images are therefore never a reliable means of evaluating $\chi$, unless numerical phase retrieval methods are used. In general, images that are recorded at different defocus values contain only partial and mixed information about the phase and amplitude of the wave, depending on the optical conditions, including defocus. As a test, we analyzed images of the virtual protein recorded at different defocus values, finding a maximum of $\langle \chi_{dig} \rangle = 0.383$ and a minimum of $\langle \chi_{dig} \rangle = -0.059$. This extreme variability results in an apparent inversion of the chirality and even a lack of reliability of its sign. This is the reason why it is challenging to use single particle analysis to identify the two orientations experimentally. For digital images of the Archimedean spiral, an apparent inversion of polarity occurs when the focus effect prevails over the phase of the object itself. As a final test, we digitally analyzed experimental images of real proteins. In many images, the 7- fold symmetry is not visible due to very low contrast. In these cases, the value of $\langle \chi_{dig} \rangle$ was erratic. However, for some cases, where the 7-fold periodicity was observed, $\langle \chi_{dig} \rangle$ was estimated to be in the range $0.16 - 0.18$,

*i.e.*, in the range obtained using digital analysis of virtual proteins. This observation is not surprising, since both $P_\rho$ and OAM are scale invariant even in digital analysis. The relatively low variability of $\langle \chi_{dig} \rangle$ in the selected small subset of digital images results from the fact that the focus is fixed. However, many other factors, such as local tilts and ice conditions, can affect $\langle \chi_{dig} \rangle$ and increase its variability.

Given the improved quality of direct $\chi$ measurement using the OAM sorter with respect to $\chi_{dig}$, the OAM sorter can be used to improve cryo electron microscopy, either alone or by combining real-space imaging and OAM-sorter space. The quantitative definition of planar chirality combined with the OAM sorter promises to improve different kinds of coherent imaging, including light. We have defined a new Hermitian operator $\chi$ and related symmetry that permits geometric planar chirality to be quantified using an OAM sorter. The sorter is able to reach a near-nominal OAM resolution of 1.1 quanta of OAM. The result of the $\chi$ measurement is not affected by imaging artifacts because it measures the amplitude and phase dichroism directly from the wavefunction, without imaging out-of-focus. We applied this to analyzed the 2D symmetry of protein in a TEM. The technique can be applied to any light or matter wave to identify and quantify planar chirality resulting from elastic scattering from an object.

**Methods**

**Virtual protein** The virtual protein and test holograms were prepared using focused ion beam milling according to a computer-generated pattern. The SiN membranes were evaporated with Au, which was removed in a circular window in the hologram region. The diameter of the circle was $20\ \mu$m for the virtual protein and $40\ \mu$m for the test holograms and spirals.

**Sorter:** The OAM sorter was implemented using electrostatic elements in an FEI Titan microscope operated at 300 kV. The hologram was inserted in the C2 diaphragm holder, while electrostatic sorter elements were inserted in the objective and selected area diffraction diaphragm holders. The spherical aberration corrector of the microscope was turned off. The probe was convergent in the specimen plane, with a convergence of 1.8 mrad. The spot size was set to 9 to ensure sufficient coherence. New elements were fabricated for these experiments, using more needles in the first sorter element for control of the phase. The lens working conditions were optimized to have a sufficiently large OAM-$P_\rho$ spectrum on the detector. Images were recorded using a Gatan K2 camera. A neural network was used to find the optimal bias for Sorter 1. The main shift alignment was performed manually.

**Practical definition of $\chi$:** A protein hologram was obtained by multslice simulation. For simplicity, only the phase of the protein was considered. For the sake of visibility, the phase was enhanced by a scaling factor of 3. However, the value of $\langle \chi \rangle$ was not affected, as explained in text. The value of $\langle \chi \rangle$ was calculated according to the formula

$$\chi \geq \frac{\sum_{k\neq 0, l\neq 0} \frac{l}{k} I_{l,k}}{\sum_{k\neq 0, l\neq 0} I_{l,k}}$$

The value of $l = 0$ was excluded because it is too dependent on the imaging condition and possible radial cutoff details.

**Supplementary material**

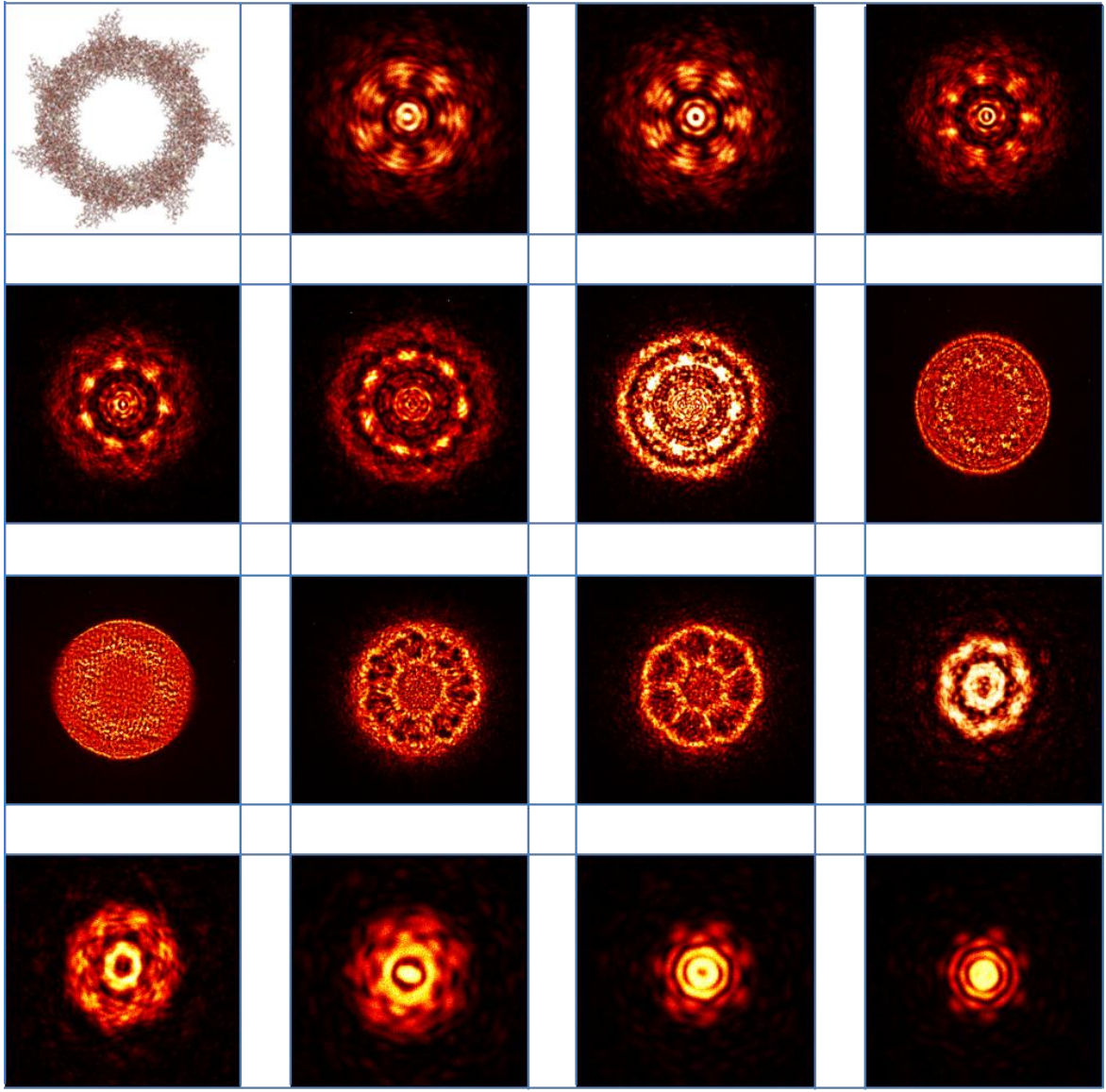

*Fig S1* *Experimental defocus series of images of a virtual protein.*

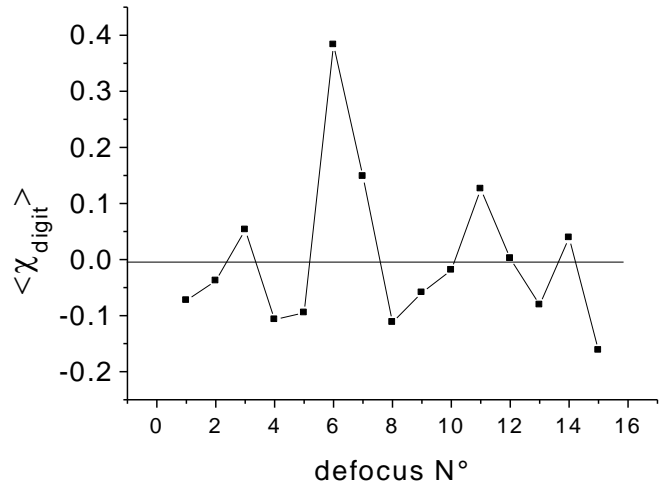

***Fig S2*** *Values of $\langle \chi_{dig} \rangle$ calculated digitally from images of the virtual protein recorded at different defocus values.*

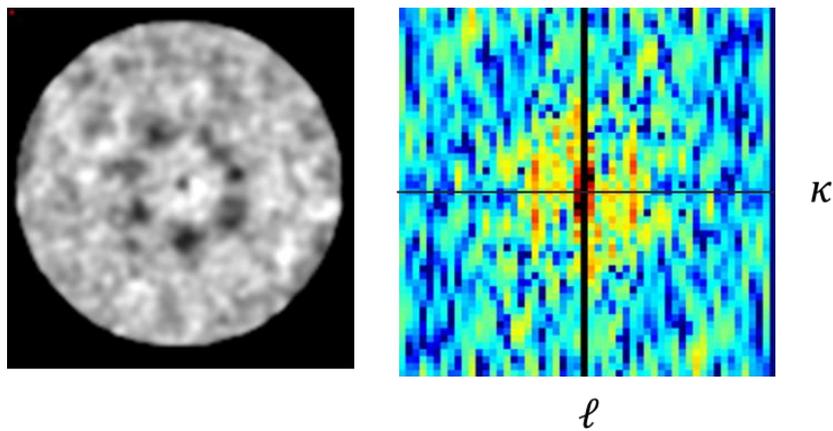

***Fig S3*** *Example of digital analysis of an experimental image of a real protein recorded using cryo electron microscopy.*

**Space representation of χ**

Since χ is a Hilbert operator, it must be possible to calculate its form in position representation. In order to obtain such a representation, we start from the eigenstates

$$u = \exp(i\ell\theta + i\kappa\rho) \,.$$

The explicit form of $\chi$ in the $\theta, \rho$ plane is

$$\chi\psi = \frac{\partial}{\partial\theta}(\Pi \otimes \psi) \,,$$

where $\otimes$ is the convolution operator (here extended to the distributional sense) and $\Pi$ is 1 if $\rho > 0$ and -1 otherwise.

Clearly, it holds that $\chi u = \frac{\ell}{\kappa} u$

[1] Noether E. Invariante Variationsprobleme. Nachr. D. König. Gesellsch. D. Wiss. Zu Göttingen, Math-Phys. Klasse, pages 235–257, 1918.

[2] Dwight E Neuenschwander. Emmy Noether's wonderful theorem. JHU Press, 2017.

[3] Wu, C. S., Ambler, E., Hayward, R. W., Hoppes, D. D., Hudson, R. P., Experimental Test of Parity Conservation in Beta Decay Physical Review. **105**, 1413–1415, (1957).

[4] Holló, Gábor "A new paradigm for animal symmetry". Interface Focus. **5**, 20150032 (2015).

[5] Finnerty, J.R. "The origins of axial patterning in the metazoa: How old is bilateral symmetry?". The International Journal of Developmental Biology. 47 (7–8): 523–9 (2003).

[6] Sir William Thomson Lord Kelvin "The Molecular Tactics of a Crystal". Clarendon Press. (1894).

[7] Tavabi a.H., Rosi P., Rotunno E., Roncaglia A., Belsito L., Frabboni S., Pozzi G, Gazzadi G.C., Lu P.H., Nijland R., Ghosh M., Tiemeijer P., Karimi E., Dunin-Borkowski R.E.:, Grillo V. Experimental Demonstration of an Electrostatic Orbital Angular Momentum Sorter for Electron Beams Physical Review Letters **126**, 094802 (2021)

[8] Schattschneider P., Rubino S., Hébert C., Rusz J., Kuneš J., Novák P., Carlino E., Fabrizioli M., Panaccione G., Rossi G., Detection of magnetic circular dichroism using a transmission electron microscope, Nature **441**, 486–488 (2006).

[9] Schnell, C. Spatially resolved isotope labeling. Nat Methods **16**, 287 (2019).

[10] Krivanek O.L:, Lovejoy T.C., Dellby N., Aoki T., Carpenter R. W., Rez P., Soignard E., Zhu J., Batson P.E., Lagos M.J., Egerton R.F. & Crozier P.A. Vibrational spectroscopy in the electron microscope Nature **514**, 209–212(2014)

[11] Li X., Haberfehlner G., Hohenester G., Stéphan O., Kothleitner G., Kociak M. Three-dimensional vectorial imaging of surface phonon polaritons Science **371**, 1364-1367 (2021)

[12] Troiani F., Rotunno E., Frabboni S., Ravelli R. B. G., Peters P. J., Karimi E., and Grillo V. Efficient molecule discrimination in electron microscopy through an optimized orbital angular momentum sorter Phys. Rev. A **102**, 043510 (2020)

[13] Zurek W.H. Decoherence, einselection, and the quantum origins of the classical Rev. Mod. Phys. **75**, 715 (2003)

[14] Grillo V, Tavabi A.H., Venturi F., Larocque H., Balboni R, Gazzadi G.C., Frabboni S, Lu P.H., Mafakheri E., Bouchard F., Dunin-Borkowski R.E., Boyd R.W., Lavery M.P.J., Padgett M.J. & Karimi E. Measuring the orbital angular momentum spectrum of an electron beam Nature Communications **8**, 15536 (2017)

[15] McMorran B.J., Harvey T.R., Lavery T.R., Efficient sorting of free electron orbital angular momentum, New J. Phys. **19**, 023053, (2017)

[16] G Pozzi, V Grillo, PH Lu, AH Tavabi, E Karimi, RE Dunin-Borkowski Design of electrostatic phase elements for sorting the orbital angular momentum of electrons Ultramicroscopy 208, 112861

[17] Pozzi G, Rosi P, Tavabi A.H., Karimi E., Dunin-Borkowski R.E., Grillo V. A sorter for electrons based on magnetic elements Ultramicroscopy in press https://doi.org/10.1016/j.ultramic.2021.113287

[18] Berkhout G.C.G., Lavery M.P.J., Courtial J., Beijersbergen M.W., Padgett M.W., Efficient sorting of orbital angular momentum states of light, Phys. Rev. Lett. **105,** 153601 (2010).

[19] J. Harris, V. Grillo, E. Mafakheri, G. C. Gazzadi, S.Frabboni, R. W. Boyd, and E. Karimi, Structured quantum waves, Nat. Phys. **11**, 629 (2015).

[20] B. J. McMorran, A. Agrawal, P. A. Ercius, V. Grillo, A. A. Herzing, T. R. Harvey, M. Linck, and J. S. Pierce, Origins and demonstrations of electrons with orbital angular momentum, Phil. Trans. R. Soc. A **375**, 20150434 (2017).

[21] K. Y. Bliokh, I. P. Ivanov, G. Guzzinati, L. Clark, R. Van Boxem, A. B´ech´e, R. Juchtmans, M. A. Alonso, P. Schattschneider, F. Nori, and J. Verbeeck, Theory and applications of free-electron vortex states, Phys. Rep. **690**, 1 (2017).

[22] S. M. Lloyd, M. Babiker, G. Thirunavukkarasu, and J. Yuan, Electron vortices: Beams with orbital angular momentum, Rev. Mod. Phys. 89, 035004 (2017)

[23] Hugo Larocque, Ido Kaminer, Vincenzo Grillo, Gerd Leuchs, Miles J Padgett, Robert W Boyd, Mordechai Segev, Ebrahim Karimi 'Twisted'electrons Contemporary Physics **59**,126 (2018)

[24] Berrier J.C., Davis B. L., Kennefick D., Kennefick J.D., Seigar M.S., Barrows R.S., Hartley M., Doug Shields, Misty C. Bentz, and Claud H. S. Lacy Further evidence for a supermassive black hole mass–pitch angle relation The Astrophysical Journal, **769**, 132 (2013)

[25] Nechayev S., Eismann J.S., Alaee R., Karimi E., Boyd R.W., and Banzer Kelvin's chirality of optical beams P. Phys. Rev. A **103**, L031501 (2021)

[26] Solomonson M, Setiaputra D, Makepeace KAT, Lameignere E, Petrotchenko EV, Conrady DG, Bergeron JR, Vuckovic M, DiMaio F, Borchers CH, Yip CK, Strynadka NCJ. , Structure of EspB from the ESX-1 type VII secretion system and insights into its export mechanism. Structure **23**, 571 (2015).

[27] R. Henderson, The potential and limitations of neutrons, electrons and X-rays for atomic resolution microscopy of unstained biological molecules Quart. Rev. Biophys. **28**, 171 (1995).

[28] R. Henderson, Avoiding the pitfalls of single particle cryo-electron microscopy: Einstein from noise, Proc. Natl. Acad. Sci. USA **110**, 18037 (2013).

[29] Grillo V., Gazzadi G.C., Karimi E., Mafakheri E., Boyd R.W., Frabboni S. Highly efficient electron vortex beams generated by nanofabricated phase holograms Applied Physics Letters **104**, 043109 (2014)

[30] Rotunno E., Tavabi A.H., Rosi P., Frabboni S., Tiemeijer P., Dunin-Borkowski R.E., Grillo V. Alignment of electron optical beam shaping elements using a convolutional neural network Ultramicroscopy **228**, 113338 (2021).